\documentclass[12pt,preprint]{aastex}
\usepackage{lscape}

\def\dec{{\rm decay}}
\def\cap{{\rm capture}}
\def\tot{{\rm tot}}
\def\eps{{\epsilon}}

\begin{document}
\title{Tritium as an Anthropic Probe}

\author{Andrew Gould}

\affil{
Department of Astronomy, The Ohio State University,
140 West 18th Avenue, Columbus, OH 43210\\
gould@astronomy.ohio-state.edu
}


\begin{abstract}

I show that if tritium were just 20 keV lighter relative to helium-3,
then the current deuterium burning phase of pre-main-sequence
stellar evolution would be replaced by deuterium+tritium burning.
This phase would take place at the same temperature but would last
a minimum of 4 times longer and a maximum of 8 times longer than
deuterium burning and so would yield total energies comparable
to the binding energy of solar-type pre-main-sequence
stars.  Hence, it could in principle
radically affect the proto-planetary disk, which forms at the
same epoch.  I suggest that this may be one of the most ``finely-tuned''
parameters required for intelligent life, with the mass range only
a few percent of the neutron-proton mass difference, and $10^{-5}$ of
their masses.  I suggest that the lower limit of this range is set by
the physics of disk formation and the upper limit by the statistical
properties of fundamental physics.  However, if this latter
suggestion is correct, the statistical distribution of physical 
``constants'' must be a power-law rather than an exponential.
I also suggest a deep connection between fundamental physics and the search
for extrasolar life/intelligence.

\end{abstract}

\keywords{early universe --- stars: pre-main sequence --- planetary systems}

{\section{Anthropic Principle: the Long View}
\label{sec:intro}}

While the standard model of particle physics and cosmology is
extremely successful in
describing phenomena, it has of order 30 parameters whose values
are increasingly well measured but with two possible
exceptions\footnote{Inflation \citep{guth81} predicted in advance
of observations that the curvature of the Universe is small
$k\sim 0$, and this has been confirmed.  Imprinting of quantum
fluctuations is also a plausible explanation for the observed
spectral index of fluctuations $n=1$.}
remain completely unexplained.
One conjecture is that there is some fundamental theory in which
all parameters are simple numbers, but through complex interactions
that will one day be reflected in equally complex calculations,
the mass of the down quark, the charge on the electron, etc., take on
their observed (seemingly meaningless) values.  A second conjecture
is that these physical constants assumed random values
when very high-energy fields froze out in the early Universe.
At present, there is no way to choose between these conjectures.

The second conjecture, however, has an interesting variant that is
at least subject to empirical investigation, if not decision.
If there are many universes, and each is endowed with random parameters,
then most of these universes will fail to produce intelligent life.
Thus speculation about the origin of physical constants is restricted
to our Universe, and to other universes whose
parameters are consistent with intelligent
life.  This variant generally goes under the heading
``Anthropic Principle''.

It is not difficult to come up with changes in physical parameters that
would gravely limit the prospects for intelligent life.  I will briefly
explore one example that is relevant to this paper.  The neutron-proton
mass difference is $\Delta M_1 = 1.293\,$MeV $(\hbar=c=1)$, which
results from slight differences in the up and down quark masses, as
well as the different charges of these particles.  Let us suppose that
all other physical constants (such as the electron mass $m_e$) were the 
same,
but $\Delta M_1=0$.  This case
illustrates both the promise and the challenges of this entire line
of investigation.  The universe would contain no hydrogen, other
than trace quantities of deuterium and tritium left over from 
Big Bang nucleosynthesis (BBN).
This would pose three severe problems for intelligent life:
no hydrogen-burning stars, no water, no organic chemistry.

Now, assuming that stars formed at all, there would still be a 
helium-burning
main sequence, which burned much hotter (and so faster) than our hydrogen
main sequence.  But, at its very bottom, there would be long-lived stars
(just as there are very long-lived stars at the bottom of our main 
sequence).
However, such stars would correspond to an extremely narrow mass range
and so would be very rare.  Now, whether such rarity is a serious
obstacle to intelligent life depends on how rare intelligent life is.
The frequency of intelligent life is a purely empirical question,
about which we happen to have at the moment extremely few data.  But
within about 1 Myr, we will have fully surveyed our Galaxy, finding
not only the intelligent life that is broadcasting radio waves, but
also that which is hiding in caves (if that turns out to be
the most intelligent thing to do).  So the answer to this question
will be known.  The problem of water looks more serious than 
hydrogen-burning
stars.  We do have evidence from Earth that life without water is at least
extremely difficult.  However, some people speculate that it may
be possible based on methane, etc.  Again, this is an empirical question,
which we will solve by direct investigation of other worlds.

Now, organic chemistry is another order of problem.  By definition,
``organic chemistry'' refers to carbon, but that is only because
chemists takes ``hydrogen'' for granted.  The previously mentioned
methane-based life (or ammonia-based life) still requires hydrogen.
This seems to be an absolute show stopper, though again we can investigate
this on other planets that lack hydrogen.

This brief summary shows that it is not trivial to determine that
some condition is required for intelligent life.  However, we can at this 
point
already develop plausible conjectures for what is required, and we will
eventually be able to address these conjectures empirically.

To have identified one value of one parameter ($\Delta M_1=0$) that
would exclude life hardly proves that this parameter is ``fine-tuned''.
``Fine-tuning'' is critical to the program of the Anthropic Principle.
If it could be shown that several parameters had very narrow intervals
within which intelligent life was possible (or at least was not much
less probable than in our Universe; e.g., the above example of very
rare long-lived stars), then we would have to conclude one of three 
things:
1) we are very lucky, 2) God made it that way, 3) there are many universes
with many different parameter sets.

However, we are very far from having to make such a choice.  Part of the
problem is lack of technology and/or funding for missions to exo-planets.
But a bigger problem is that we do not usually analyze systematically
how our Universe
would be different if some parameter or other was varied.  Of course,
this is routine when the parameter's measurement is uncertain, but
we do not routinely ask why the Universe has its observed
features, by going far
enough beyond the uncertainties in parameters to figure out at what
point it would be different.

Of course, there has been substantial work to identify
parameters (or parameter combinations) that are clear show stoppers
in order make ``conservative'' arguments that at least some observed
parameters lie in a narrow range that permits life.  See for example,
\citet{livio89} or \citet{hogan06}.
these below.  But here I mean something different.  If there are
indeed many universes with randomly assigned ``fundamental'' parameters,
then a huge fraction will be saddled with one or more characteristics
that are catastrophically
incompatible with intelligent life.  But of those that remain
some will be more intelligent-life friendly than others, because
more of their parameters lie close to optimal.  And depending on
how rare intelligent life is, some universes may prove 
intellectually-sterile
as a result of an accumulation of small nicks rather than a few
huge blows.

This approach to the Anthropic Principle is much messier, much more
challenging than the standard one exactly because it requires a much
deeper understanding of our own Universe.  But it also
points to a potentially deep connection between two currently disconnected
research domains: search for extra-solar life and fundamental physics.

A large number of theorists of various stripes, mostly particle
physicists and cosmologists regard the search for other worlds,
including their possible life forms, as being of no intrinsic
scientific interest and possibly not even science.  But it may be
that these investigations provide one of the best clues to physics
at the highest energies.

To make this clearer, let us relax the extreme example above,
and permit positive values of $\Delta M_1$, but not much heavier
than the electron mass.  In this regime, the neutron decay rate
scales (in analogy to Equation (\ref{eqn:den2}))
as $(\Delta M_1 - m_e)^4$ and so can for present purposes be
ignored.  Then the proton fraction $f_p$ coming out of
BBN would be
\begin{equation}
f_p = \tanh(\Delta M_1/2 T_{fo})\simeq \Delta M_1/2 T_{fo},
\label{eqn:fp}
\end{equation}
where $T_{fo}=0.72\,$MeV is the neutron freeze-out temperature.
Now, for $\Delta M_1<m_e$, the hydrogen atom is unstable to
electron capture, which is the type of catastrophe that is
often considered (e.g., \citealt{hogan06}).  
But suppose $\Delta M_1=m_e$ or just above.
Then $f_p=35\%$.  The most elementary questions about such a universe
remain untabulated, such as what fraction of stars would live long
enough (say 4 Gyr) to nurture intelligent life, or would planets
orbiting such stars at distances suitable for liquid water
be tidally locked?  And, ipso facto, nothing is tabulated about
more difficult questions under such conditions, such as water delivery
to rocky planets.  Of course, part of the reason these questions
remain unanswered (actually unasked) is that the answers would
not lead to any immediate conclusions, since we do not yet have
the data necessary to determine whether tidal locking is
adverse to life, or whether intelligent life is so rare that major
reductions in habitability could plausibly reduce the chance
for intelligent life in our Universe to much less than unity.
But part of the reason is simply lack of imagination and
narrowness of vision about the nature of the scientific enterprise:
we can gain just as much insight into the nature of our Universe
by considering paths not taken \citep{frost20} as by measuring the details of
observable parameters.

Here I show that $\Delta M_3$,
the mass difference between tritium (T) and helium-3
($^3$He) may provide an even more fine-tuned lower limit on the
mass of the neutron.  This difference
\begin{equation}
\Delta M_3 = 2.00\times 10^{-5}\, {\rm AMU} = 18.6\,{\rm keV}
\label{eqn:m3}
\end{equation}
(between dressed atoms, not bare nuclei) is almost 2 orders of
magnitude smaller than $\Delta M_1$ (and 5 orders of magnitude
below the proton mass).  Specifically, if $\Delta M_3$ were even
20 keV smaller (i.e., negative by at least a few keV) then
pre-main-sequence stellar evolution, and so disk formation, would
be significantly impacted.  Determining whether such impact has important
implications for intelligent life is well beyond the scope of this
paper: my main objective is to point out that the issue exists.

Even if it turns out that this limit is important, it is a one-sided
limit and therefore not, in itself, a case of ``fine-tuning''.  I
discuss why such 1-sided limits may be automatically two-sided and
therefore relevant to the issue of fine tuning.

{\section{Light-quark Anthropics: A Different Approach}
\label{sec:lightquark}}

There are two light quarks, the down and up, and so potentially
two Anthropic constraints on their masses.  In practice, the
physical arguments relate to the sum and difference of these masses,
which impact nuclear physics in qualitatively different ways.
The sum affects the mass of the pion, which mediates
inter-nucleon forces.  The difference goes directly into $\Delta M_1$,
the neutron-proton mass difference, which helps determine nuclear
stability and also nucleosynthesis.

Within the standard model, the quarks obtain their masses from 
the Higgs (characterized by a dimensionful mass) via Yukowa coupling
factors (which are dimensionless).  For reasons that do not concern
us here, some particle physicists consider that if there were
one particular parameter that was randomly assigned, it would be the
Higgs mass.  That is, they hold out hope that the dimensionless
Yukowa couplings might be fully predictable.  Thus, they focus
on a single-parameter variation of quark masses, which is
parameterized by (confusingly) $w$.  All quark masses rise or fall
together by $w^{1/2}$, with $w=1$ characterizing our Universe.

This has the advantage that it permits all constraints to be put
on a single axis, and in particular it allows one to derive both
upper and lower bounds for $w$ by combining arguments based on
intra-nuclear forces and nuclear stability, respectively.

However, I would advocate the opposite approach: sticking as
close to the data as possible.  As I have tried to argue, it is already
quite difficult to figure out the implications of changing a single
observable parameter, such as $\Delta M_1$ for the prospects of
intelligent life.  And the connections of these parameters to some
fundamental theory remain, at this point, quite speculative.  Thus,
it is best to try to develop a solid base of knowledge of individual
parameters, and then to evaluate the impacts on a range of theories
according to the predictions that those theories make about the
parameters.  Then when grouping various arguments together for
confrontation with theories, this should still be done on a
semi-empirical basis.  That is, arguments related to the neutron-proton
mass difference ($\Delta M_1$) and tritium-helium-3 mass difference
($\Delta M_3$) are almost certainly both primarily rooted in
the down-up quark mass difference and so can be grouped together.
Thus, in evaluating their combined impact, one can keep in mind
that they all probably move together, if not perfectly in tandem.
However, in keeping with the ``long view'' advocated in the previous
section, I think it is premature to put all arguments, even those
related to sums versus differences of light-quark masses, on a common
scale.

{\section{Electron Capture by Helium-3}
\label{sec:ec}}

If helium-3 were lighter than tritium (but by less than $2\,m_e$), then
it would decay into tritium
by electron capture (EC), as for example occurs with beryllium-7 and
argon-37.  Before changing universes, I first evaluate the rate of
EC for helium-3 in our universe, which is energetically forbidden
for bound electrons, but possible for energetic
free electrons.  Consider a flux $F$ of electrons with kinetic
energy $m_e v^2/2 = E+\Delta M_3$, so that the total kinetic energy
of the final state is $E$.  Then the rate of capture is
\begin{equation}
\Gamma_\cap = F A V_\cap;
\quad
V_\cap\equiv\int d^3 p_\nu \int d^3 p_{He}
\delta^4([{\bf p}_\tot,E_\tot] - [0,E])
\label{eqn:cap}
\end{equation}
where $A$ contains all the low-energy physics of the interaction,
$V_\cap$ is the phase-space integral, and the calculation
is done in the center-of-mass frame.  This quantity can be related to
the {\it observed} decay rate of tritium, $\Gamma_\dec=\ln 2/t_{1/2}$,
($t_{1/2} = 12.3\,$yr):
\begin{equation}
\Gamma_\dec = A V_\dec;
\quad
V_\dec\equiv\int d^3 p_e \int d^3 p_\nu \int d^3 p_T
\delta^4([{\bf p}_\tot,E_\tot] - [0,\Delta M_3])
\label{eqn:dec}
\end{equation}
where $A$ is identical to the same term in the previous equation.
Hence, the ratio of these rates is
\begin{equation}
R\equiv{\Gamma_\cap\over\Gamma_\dec}
= F {V_\cap\over V_\dec}
\label{eqn:ratio}
\end{equation}

Noting that
$m_\nu\ll\Delta M_3 <m_e\ll m_p$, we can easily evaluate the
numerator and denominator:
\begin{equation}
V_\cap = 4\pi E^2
\label{eqn:num}
\end{equation}
\begin{equation}
V_\dec
=
(4\pi)^2\int_0^{\Delta M_3} d\eps\,\eps^2 [(\Delta M_3 + m_e - \eps)^2- 
m_e^2]
\label{eqn:den}
\end{equation}
\begin{equation}
=(4\pi)^2(\Delta M_3)^4\biggl({5\over 6}m_e + {\Delta M_3\over 30}\biggr)
\rightarrow {5\over 6}(4\pi)^2 m_e(\Delta M_3)^4,
\label{eqn:den2}
\end{equation}
where I have dropped a term of order 0.1\% in the last step, for
simplicity.  Hence,
\begin{equation}
R = {6/5\over 4\pi} \biggl({E\over \Delta M_3}\biggr)^2\,
{F\over m_e(\Delta M_3)^2}
\label{eqn:ratio2}
\end{equation}

Let us now consider an alternate universe in which tritium is lighter than
helium-3 by $\Delta M_3'$ (instead of heavier by $\Delta M_3$
as in our Universe).  That is, $\Delta M_3'>0$ by convention.

And let us first examine the case of a flux of electrons of density $n_e$
and temperature $T\ll \Delta M_3'$.  We may then approximate
$E\simeq\Delta M_3'$.
Since $F=\sqrt{8T/\pi m_e}n_e$,
we can evaluate $R'$, the ratio of electron capture by helium-3
in the other universe to the tritium decay rate in our Universe
\begin{equation}
R' = {3\over 5}\sqrt{2\over\pi^3}\biggl({\Delta M_3'\over \Delta 
M_3}\biggr)^2
{n_e\over (\Delta M_3 m_e)^{3/2}}\sqrt{T\over \Delta M_3} =
1.26\times 10^{-4}n_{26}\biggl({\Delta M_3'\over \Delta M_3}\biggr)^2\,
\sqrt{T\over \Delta M_3}
\label{eqn:ratioprime}
\end{equation}
where $n_{26}$ is the electron density normalized to $10^{26}\,\rm 
cm^{-3}$.
Hence, for example, in the center of the Sun, where $n_{26}\sim 1$ and
$T/M_3\sim 0.07$, the half life of helium-3 by EC would be about 0.4 Myr
if $\Delta M_3' = \Delta M_3$.  I return to this point below.

We can also estimate the capture rate of an electron in a bound orbit
of (for simplicity) a singly-ionized helium-3 nucleus.  To do so,
I approximate the flux as
\begin{equation}
F \simeq  {\beta\over \pi a^3} = {m_e^3 (Z\alpha)^4\over \pi}
\label{eqn:fbound}
\end{equation}
where $\beta = Z\alpha$ is the bound electron velocity, $Z=2$ is
the atomic number, and $a=(m_e Z\alpha)^{-1}$ is the generalized Bohr
radius.
This implies
\begin{equation}
R' = {6/5\over \pi^2} \biggl({\Delta M_3'\over \Delta M_3}\biggr)^2\,
\biggl({E_b\over \Delta M_3}\biggr)^2 \sim 9\times 10^{-7}
\biggl({\Delta M_3'\over \Delta M_3}\biggr)^2
\label{eqn:rbound}
\end{equation}
where $E_b=54\,$eV is the binding energy of singly-ionized helium.
That is, for $\Delta M_3'=\Delta M_3$, the
half-life of helium-3 (bound to at least one electron) would be
$\tau_{1/2}=14\,$Myr.

{\section{Alternate Universe History}
\label{sec:altuni}}

{\subsection{From BBN to Dark Ages}
\label{sec:bbn}}

We are now in a position to trace the role of tritium in an alternate 
universe
with, say $\Delta M_3' = \Delta M_3$,
from the Big Bang onward.  Its Big Bang itself would be extremely
similar to ours because the difference in $\Delta M_3$
(i.e., $\Delta M_3 +\Delta M_3'=37\,$keV), is very
small compared to all relevant energies during that epoch.  After
the Big Bang, however, the tritium in our Universe decays into helium-3
while in this universe it would not.  But neither would the helium-3
decay into tritium: this is energetically forbidden.  Nor would it 
initially
capture electrons, since the helium-3 is fully ionized and the ambient
density of electrons is too low.  Nevertheless, after helium 
recombination,
the helium-3 nuclei would capture electrons on 14 Myr timescales, so that 
by
the time of star formation all of it would be converted to tritium.
Hence, at this time, this universe would look very much like our Universe,
except that all the helium-3 would be replaced by tritium.

{\subsection{Star Formation}
\label{sec:SF}}

However, the process of star formation would look very different.
The most important difference is that tritium burns to helium-4 at
approximately the same temperature that deuterium burns to helium-3.  This
means that the phase of ``heavy hydrogen burning'', which we now designate as
``deuterium burning''  would be generalized to
``deuterium and tritium burning''.
Let us initially assume that none of the helium-3 that is produced by
deuterium burning is converted to tritium.  Then, since there is
about 2/3 as much tritium as deuterium, and since burning tritium
yields about 4 times as much energy as burning deuterium to helium-3,
the total energy generated during this phase would increase by a
factor $1 + 4\times (2/3)\sim 4$ relative to our Universe.
Hence, the phase of
``heavy-hydrogen burning'' would last about 4 times longer
than the current ``deuterium burning''.

Moreover, we can expect that a substantial fraction of the newly
generated helium-3 would convert to tritium during this epoch.  Paradoxically,
it is too hot for electron capture onto helium-3 in the stellar core
where it is created.  The temperature required for deuterium burning
is about $T=0.1\,$keV, at which point helium is almost fully ionized.
And the ambient electron density,
$n_{26}\sim 0.01$ implies that capture from free electrons is also
too slow.  See Equation (\ref{eqn:ratioprime}).

However, pre-main-sequence stars are fully convective.
Hence, just as deuterium (and now tritium) are fed to the core from
the entire star, the helium-3 ash from deuterium burning is sent
to the cooler outer portions of the star where it can capture
electrons into orbit, and from there into the nucleus.  If
all this helium-3 were converted, it would add another factor of two
to the energy output, and so to the timescale, i.e., 8 times longer than
at present.

To understand the importance of this effect, it is useful to
compare the energy generated by deuterium burning
to the gravitational potential energy of the star at this stage,
when it typically has a radius that is 5 times larger than on the
main sequence.  Specifically for solar-type stars in our Universe,
the ratio of energy released in deuterium burning to gravitational
potential energy is approximately
\begin{equation}
W_D = {(N_D/N_H)(M_H/M)(E_D/m_p)\over GM/R} = 0.34
\label{eqn:wd}
\end{equation}
where $N_D/N_H=3\times 10^{-5}$ is the deuterium-to-hydrogen ratio,
$M_H/M=0.75$ is the fraction of mass in hydrogen, and
$E_D=5.6\,$MeV is the energy released from deuterium burning.

But in the other universe, the corresponding quantity, $W_{D+T}$,
would lie in
the range $1.4< W_{D+T}< 2.8$.  Thus, this period of stellar evolution, 
when
gravitational contraction is temporarily halted by the onset of
nuclear fusion, would go from a ``bump in the road'' to an
independent phase.  Another way to state this is that the phase of
heavy-hydrogen burning would generate 3--6 Myr-$L_\odot$ of total 
energy, rather than 0.7 Myr-$L_\odot$ at present.
Since the protoplanetary disk is taking
shape during near the time of heavy-hydrogen burning and on Myr timescales, 
it is at least possible that this
change would radically impact disk evolution, and possibly the
formation of planets.

{\subsection{Stellar Evolution}
\label{sec:evolution}}

After the deuterium and tritium were exhausted, the star would
continue its contraction until hydrogen fusion was ignited.
As mentioned following Equation (\ref{eqn:ratioprime}), the $p-p$
chain would be significantly altered: the path from helium-3
to helium-4 would primarily go through EC on helium-3 to
produce tritium, which would then immediately burn to helium-4.
However, this substitution would have only minor practical consequences,
at least in stars like the Sun.  In the Sun, virtually all the
neutrons are produced in $p-p$ fusion.  But in this other universe,
almost half would be produced by EC capture.  Since $p-p$ burning
is the main bottleneck in the Sun, the central temperature would
be reduced to a level that halved the rate of this process.  However,
since $p-p$ burning is very temperature sensitive, the actual
temperature reduction (and so the impact on stellar structure) would
be small.

It is true that brown dwarfs would burn fuel for eight times longer
than in our Universe, but it is difficult to imagine major impacts
of this fact on the course of intelligent life.

{\section{Other Values of $\Delta M_3'$}
\label{sec:altvals}}

We should now consider values of $\Delta M_3'$, other than being exactly
equal to $\Delta M_3$.  Clearly, if $\Delta M_3'>\Delta M_3$, then
the entire story is similar except that helium-3 is more quickly
converted to tritium, so that deuterium+tritium burning moves rapidly
toward lasting the maximum of the allowed range, i.e.,
8 times longer than in our Universe.

For smaller $\Delta M_3'$, the opposite is true: the timescale
for EC increases so that deuterium+tritium burning moves rapidly
toward the minimum of the allowed range, i.e.,
4 times longer than in our Universe.

At very small values,
$\Delta M_3'\la 0.1\,\Delta M_3$, the half-life of helium-3 gets so long,
$\tau_{1/2}\ga 1\,$Gyr, that the BBN endowment of helium-3 does not
convert to tritium prior to star formation (or before it is shut off
by helium reionization).  Now, because of their 
fixation on our own Universe, cosmologists almost never even plot
the original tritium endowment versus $\eta_b$,
the baryon-to-photon ratio.  By chance however,
the one example given by \citet{wagoner73} is for ``$h_0=10^{-4.5}$'',
which one finds after some algebra is equivalent to
$\eta_b=4.5\times 10^{-10}$, i.e., very close to the modern value
$\eta_b=6.2\times 10^{-10}$.  His Figure 2 shows that after BBN, but
before tritium decays, the ratio of tritium to helium-3 is only about
10\%.  Therefore, if $\Delta M_3'\la 2\,$keV, so that helium-3 fails
to convert, then tritium burning plays only a minor role in star 
formation.

{\section{Two-sided limit?}
\label{sec:twodied}}

Even if it is eventually shown that disk formation would be radically
impacted by tritium burning, this would not by itself be an example
of ``fine tuning''.  There would just be a one-sided requirement that
tritium be heavier than helium-3 (or at least no more than 2 keV lighter),
which does not appear very constraining.  It would then be a bit curious
that our Universe had just barely met this threshold, but not
in itself clear evidence for fine-tuning.

Now, the observed value of $\Delta M_3$ is just one of many
relations in nuclear physics that are directly impacted by
$\Delta M_1$, the neutron-proton mass difference, which in turn
is a direct consequence of
$\Delta M_{1/3} = M_{\rm down} - M_{\rm up}$, the
mass difference between the down and up quarks.  It is possible
that if these more fundamental quantities were larger by, say,
a few tens of keV, then there would be some other adverse
consequence for intelligent life.  This possibility should be
explored.

However, another possibility is that when it is understood how the
up and down quark masses are randomly assigned, it will be found that
high values of $\Delta M_{1/3}$ are
suppressed.

In our Universe, $\Delta M_{1/3}=2.8\,$MeV.
It is this difference that makes the neutron heavier than the proton.
If the up an down quark masses were equal, $\Delta M_{1/3}=0$, then
the neutron would be lighter than the proton by $\Delta M_1=-1.5\,$MeV
because is has less electrostatic energy.
As discussed in Section~\ref{sec:intro}, such universes (or even universes
with $\Delta M_1=0$) would not contain life.  Nevertheless, it
may be that $\Delta M_{1/3}=0$ is the ``natural value'' in the true
fundamental theory, and only a small fraction of random realizations
have values as large as that observed in our Universe.

Without knowing the fundamental theory, there is a limited amount
we can say about such suppression, but we can say that it is not
exponential.  If it were, then the fraction of universes with
$\Delta M_3$ near the observed value would be of order
$\ln P\sim -\Delta M_{1/3}/2\Delta M_3= -75$.  Now, if the
challenges posed by tritium burning to intelligent life were as severe
as those posed by absence of hydrogen, then it would be plausible
that we would inhabit one of the one in $10^{30}$ universes that
satisfied the tritium constraint.  However, given the diversity
of astrophysical phenomena, it is virtually impossible that
tritium burning could interfere with planet formation at this
level.

Therefore, if the upper bound on $\Delta M_3$ does come from
suppression of high $\Delta M_{1/3}$ in the statistical
distribution of realizations of a fundamental theory, then
this suppression must have a form that is much weaker than exponential,
for example, power-law.


\acknowledgments
I thank John Beacom, Marc Pinsonneault, and Gary Steigman for
stimulating discussions.

This work was supported in part by grant AST 1103471 from the NSF.

\clearpage

\clearpage


\end{document}